\documentclass[aps,pra]{revtex4}

\usepackage[dvips]{graphics}
\usepackage{graphicx}
\usepackage{amsfonts}
\usepackage{amssymb}
\usepackage{amsmath}

\begin{document}

\title{Soliton Dynamics in Linearly Coupled Discrete Nonlinear Schr\"odinger Equations}
\author{A. Trombettoni$^1$, H.E. Nistazakis$^2$, Z. Rapti$^3$,
D.J. Frantzeskakis$^2$,  and P.G. Kevrekidis$^4$} 
\affiliation{
$^1$ International School for Advanced Studies and Sezione INFN, Via Beirut 2/4, I-34104, Trieste, Italy \\
$^2$ Department of Physics, University of Athens, Panepistimiopolis, Zografos, Athens 15784, Greece \\
$^3$ Department of Mathematics, University of Illinois at Urbana-Champaign, Urbana, Illinois 61801-2975, USA \\
$^4$ Department of Mathematics and Statistics, University of Massachusetts, Amherst MA 01003-4515, USA }

\begin{abstract}
We study 
soliton dynamics in a system of two linearly coupled discrete nonlinear Schr\"odinger equations, 
which describe the dynamics of a two-component Bose gas, coupled by an electromagnetic field, and 
confined in a strong optical lattice. When the nonlinear coupling strengths are equal, 
we use a unitary transformation to remove the linear coupling terms, 
and show that the existing soliton solutions oscillate from one species to the other. 
When the nonlinear coupling strengths are different, 
the soliton dynamics is numerically investigated and the findings are compared 
to the results of an effective two-mode model. The case 
of two linearly coupled Ablowitz-Ladik equations is also investigated. 
\end{abstract}
\maketitle

\section{Introduction}

The study of discrete solitons is a central topic in nonlinear lattice dynamics \cite{flach98,scott99,hennig99}. 
Of particular relevance is the robustness of soliton motion against perturbations and 
the possibility of controlling the soliton dynamics. 
Accordingly, many recent studies on this subject are focused on multi-component nonlinear lattices 
(see, e.g., 
\cite{peli1,peli2} and references therein). 
In this context, the inter-component coupling has a crucial role, as it may either destabilize the soliton propagation, 
or it can be used as a resource to move the soliton from one component to the other in a controllable manner.

The interest in multi-component systems is considerably motivated by  the experimental realization and control 
of mixtures of Bose-Einstein condensates (BECs) composed either by different hyperfine states \cite{binary1,binary2,mertes07} 
(including spinor BECs \cite{spinor}), or by different species \cite{KRb,LiCs}. 
Importantly, multi-component BECs can be trapped in strong optical lattices, as in the recent experiment 
reported in Ref. \cite{minardi07}. 
The optical lattice can be created by one or more pairs of counterpropagating laser beams, 
giving rise to a periodic potential in one, two, or three spatial dimensions \cite{morsch06}. In such a case, 
the dynamics of {\it each} (single-component) BEC is typically well described by a discrete dynamical 
model, namely the discrete nonlinear Schr\"odinger equation (DNLSE) \cite{trombettoni01} - see also 
\cite{kevrekidis01,morsch06,smerzi07} for reviews and references therein. 
Denoting by $\Psi_j$ the BEC wavefunction on site $j$ of the optical lattice, this model reads
\begin{equation}
\label{DNLS}
i  \hbar \frac{\partial \Psi_j}{\partial t} = - K
(\Psi_{j-1} + \Psi_{j+1}) + u \mid \Psi_j \mid ^2 \Psi_j +
V_j \Psi_{j},
\end{equation}
where $K$ is the tunneling rate between neighboring sites, $u$ is the nonlinear coefficient proportional 
to the s-wave scattering length, and $V_j$ accounts for an external potential 
that may be superimposed to the lattice. 
Notice that although Eq. (\ref{DNLS}) has been written for a one-dimensional ($1D$) setting, 
a generalization to higher-dimensions is straightforward. The DNLSE is a typical example of a discrete dynamical system, 
whose properties have been intensively studied \cite{hennig99,kevrekidis01,ablowitz04}. The interest for the DNLSE 
model arises from the fact that it has been successfully used to describe the dynamical 
properties of several systems 
(including, e.g., arrays of coupled optical waveguides \cite{eisenberg98}). 
In $1D$, the DNLSE is not integrable \cite{ablowitz04}; 
however, and as concerns the homogeneous case ($V_j=0$), 
soliton-like wavepackets exist and can propagate for a long time as stable objects, as it can be shown, e.g., 
by variational approaches \cite{malomed96,aceves96}.
Furthermore, the dynamics of such traveling pulses has been investigated in detail in the literature 
\cite{duncan93,flach99,gomez04,cuevas,oxtoby}. 

Based on the above discussion, here we consider the dynamics of a multicomponent BEC in an optical lattice  
described by a system of coupled DNLSEs, the different BEC components being different hyperfine levels 
\cite{mandel03}. Since the atoms of different components 
interact with each other, one has a {\em nonlinear} coupling between the different DNLSE; 
furthermore, one can induce a {\em linear} Rabi coupling by coupling the different hyperfine levels 
by an electromagnetic field \cite{ketterle,emergent}. 

The prototypical system we consider is a two-component Bose gas in an external trapping potential: 
typically the condensates are different Zeeman levels of alkali atoms like $^{87}$Rb. 
Experiments with a two-component $^{87}$Rb condensate use atom states customarily denoted by 
$\vert 1 \rangle$ and $\vert 2 \rangle$; in particular,  
the states can be $\vert F=2,m_F=1 \rangle$ and $\vert 2,2 \rangle$,  
like, e.g., in \cite{smerzi03}, or $\vert 1,-1 \rangle$ and 
$\vert 2,1 \rangle$, like, e.g., in \cite{williams00} (see also the recent work \cite{mertes07}). 
In general, the condensates $\vert 1 \rangle$ and $\vert 2 \rangle$ have 
different magnetic moments: then in a magnetic trap they can be 
subjected to different magnetic potentials, eventually 
centered at different positions and having the same frequencies (like in the setup described 
in \cite{williams00}) or different frequencies \cite{smerzi03}; in the latter work,  
the ratio of the frequencies of $V_2$ and $V_1$ is $\sqrt{2}$. 
It is also possible to add a periodic potential acting on the two-component 
Bose gas \cite{mandel03}. 

The two Zeeman states $\vert 1 \rangle$ and $\vert 2 \rangle$ can be coupled by an electromagnetic field 
with frequency $\omega_{\rm ext}$ and strength characterized by the 
Rabi frequency $\Omega_R$. A discussion of (and references on) the experimental 
manipulation of multicomponent Bose gases are in \cite{ketterle,emergent}. 

In the case of a binary BEC mixture confined in a one-dimensional lattice 
the system of the two coupled DNLSEs takes the form
\begin{eqnarray}
i \hbar \frac{\partial \psi_{(1)j}}{\partial t}=-K_1 \left( \psi_{(1)j-1} + \psi_{(1)j+1} \right)+
u_{11}\vert \psi_{(1)j}\vert ^{2} \psi_{(1)j} 
+u_{12}\vert \psi_{(2)j}\vert ^{2} \psi_{(1)j} +V_j \psi_{(1)j}+\Omega(t) \psi_{(2)j},
\label{lceqa} \\
i \hbar \frac{\partial \psi_{(2)j}}{\partial t}=-K_2 \left( \psi_{(2)j-1} + \psi_{(2)j+1} \right) +
u_{12}\vert \psi_{(1)j}\vert ^{2} \psi_{(2)j} 
+u_{22}\vert \psi_{(2)j}\vert ^{2} \psi_{(2)j} +V_j \psi_{(2)j}+\Omega(t) \psi_{(1)j}.
\label{lceqb}
\end{eqnarray}
In Eqs. (\ref{lceqa})-(\ref{lceqb}), $\psi_{(\alpha)j}$ denotes the wavefunction of the condensate $\alpha$ 
($\alpha=1,2$) on site $j$, $\Omega(t)$ is proportional to the Rabi frequency (i.e., the strength of the 
electromagnetic field), the nonlinear coefficients $u_{\alpha \beta}$ are proportional to 
the scattering lengths $a_{\alpha \beta}$ for species $\alpha$ and species 
$\beta$, 
accounting for intra- ($\alpha=\beta$) and inter- ($\alpha \neq \beta$) species interactions. 
In Eqs. (\ref{lceqa})-(\ref{lceqb}) we also assumed that the external potential superimposed 
to the optical lattice is equal for both species. 
Furthermore, $K_\alpha$ is the tunneling rate between nearest-neighbour sites for particles of the condensate 
$\alpha$; 
for a two-component condensate, when the periodic potential is the same for both the hyperfine levels 
one has $K_1=K_2$, otherwise one has $K_1 \neq K_2$ \cite{nota}.  

In this work we are interested in studying the dynamics of the system, which is prepared 
with a soliton initially present in a single component, 
after the linear coupling $\Omega$ is turned on. Our presentation is organized as follows. 
First, in Section II we show that for $K_1=K_2$ it is possible to perform 
a unitary transformation to remove the linear coupling, analogous to the one valid for two linearly coupled 
continuous Gross-Pitaevskii equations \cite{decon} (see also earlier relevant work in the 
context of nonlinear optics \cite{trans,bradley}); 
one can then use the results valid for a single-component 
DNLSE to determine the soliton dynamics. The case $K_1 \neq K_2$ is numerically studied, 
and our findings are compared in Section III with the results of a simplified two-mode model. 
In Section IV, we consider two linearly coupled Ablowitz-Ladik equations (which is an integrable 
variant of the DNLSE \cite{ablowitz76}) showing results similar to those obtained for two coupled DNLSEs. Finally, in section V, we 
briefly summarize our findings and present some interesting directions for future work.

\section{Equal Tunneling Rates}

\subsection{General Lattice}

In this Section we study the dynamics of a soliton initially present in one of the components, when the linear coupling 
with the other component is turned on. Although we are interested in solitons in $1D$ coupled DNLSEs, 
to show the generality of our approach we write the Eqs. (\ref{lceqa})-(\ref{lceqb}) on a general lattice as
\begin{eqnarray}
i \hbar \frac{\partial \psi_{(1)j}}{\partial t}=-K_1 \sum_{n(j)} \psi_{(1)n} +
u_{11}\vert \psi_{(1)j}\vert ^{2} \psi_{(1)j} +
u_{12}\vert \psi_{(2)j}\vert ^{2} \psi_{(1)j} +V_j \psi_{(1)j}+\Omega(t) \psi_{(2)j},
\label{lceqa-gen} \\
i \hbar \frac{\partial \psi_{(2)j}}{\partial t}=-K_2 \sum_{n(j)} \psi_{(2)n}  +
u_{12}\vert \psi_{(1)j}\vert ^{2} \psi_{(2)j} 
+u_{22}\vert \psi_{(2)j}\vert ^{2} \psi_{(2)j} +V_j \psi_{(2)j}+\Omega(t) \psi_{(1)j}, 
\label{lceqb-gen}
\end{eqnarray}
where the sum $\sum_{n(j)}$ is on the neighbors $n$ of the site $j$. We will also consider equal 
interaction-strengths ($u_{11}=u_{22}=u_{12}=u$), as it is almost the case, e.g., for the two-component $^{87}$Rb BECs 
studied in \cite{williams00,smerzi03}; 
in fact, in Ref. \cite{williams00}, the ratios $u_{11}:u_{12}:u_{22} = 0.97:1.00:1.03$ were used, while 
in Ref. \cite{smerzi03} $u_{11}:u_{12}:u_{22} = 1.00:1.00:0.97$. 

When $K_1=K_2$ Eqs. (\ref{lceqa-gen})-(\ref{lceqb-gen}) can be rewritten in a more compact form as:
\begin{equation}
\label{eqn:stark:dnls}
i \hbar \frac{\partial {\psi_j}}{\partial t}  =
{\cal T} \psi_j+({\psi_j}^\dagger G \psi_j)
\psi_j+V_j \psi_{j}+\Omega(t) P \psi_j,
\end{equation}
where
\begin{equation}
\label{psi:G:dnls}
\psi_j=\left( 
\begin{array}{c}
\psi_{(1)j}\\
\psi_{(2)j}
\end{array}
\right), ~~
G=u\left(
\begin{array}{cc}
1&0\\
0&1
\end{array}
\right), ~~ 
P=\left(
\begin{array}{cc}
0&1\\
1&0
\end{array}
\right),
\end{equation}
and 
\begin{equation}
\label{T:dnls}
{\cal T} \psi_j \equiv -K \sum_{n(j)} \psi_n.
\end{equation}

Then, similarly to the case of the continuum counterpart of our model \cite{decon,trans,bradley}, 
%
%
we introduce the spinor field $\phi_i$ through the relation
\begin{eqnarray}
\psi_i=U(t) \phi_i,
\label{stark1:dnls}
\end{eqnarray}
where $U(t)$ is given by 
\begin{equation}
\label{unit:dnls}
U(t)=\exp{\left[ -i P {\cal I}(t)  \right]} = \left(
\begin{array}{cc}
\cos{{\cal I}(t)}& -i \sin{{\cal I}(t)}\\
-i \sin{{\cal I}(t)} & \cos{{\cal I}(t)}
\end{array}
\right),
\end{equation}
with ${{\cal I}(t)}=(1/\hbar) \int_0^t\Omega(t') dt'$, 
and obtain, 
\begin{equation}
i \hbar \frac{\partial {\phi_j}}{\partial t}  =
{\cal T}\phi_j+({\phi_j}^\dagger G \phi_j)
\phi_j+V_j \phi_{j},
\label{eqn:dnls}
\end{equation}
which corresponds to Eq. (\ref{eqn:stark:dnls}) {\it without} the Rabi term proportional to $\Omega(t) P$. This 
transformation was originally introduced in Ref. \cite{trans} (see also \cite{bradley}), where it was used 
to eliminate a constant linear coupling term by a change of polarization basis in a set of two coupled {\it continuous} 
NLS equations modeling optical pulse propagation in birefringent optical fibers, and later was used in the context 
of binary BECs in Ref. \cite{decon}. 
It is important to stress that the above transformation (\ref{stark1:dnls})-(\ref{unit:dnls}) 
eliminates the Rabi term only when $G$ and $P$ commute; in other words, the transformation is only possible in the 
so-called Manakov case \cite{manakov}, where all interaction strengths are equal, i.e., $u_{11}=u_{22}=u_{12}$. 
In the case of nonequal intra- and inter-species interaction strengths, i.e., $u_{11}=u_{22} \ne u_{12}$,  
the transformation can still be applied but the final equation contains a more involved nonlinear interaction term. 
This physically relevant situation was recently considered in Ref. \cite{cont}, where 
the latter calculation was carried out for the continuous case.  

Let us now consider a soliton solution of Eq. (\ref{eqn:stark:dnls}) with the the linear coupling turned off ($\Omega(t)=0$) 
with initial condition
\begin{equation}
\psi_j(t=0)=\left( 
\begin{array}{c}
\Psi_{j}(t=0)\\
0
\end{array}
\right),
\label{t-0}
\end{equation}
i.e., all particles are initially in the first component. Here, $\Psi_j(t)$ denotes the soliton solution of the 
single component DNLSE for the condensate $1$, which we assume to be known. If at time $t=t_0$ the linear 
coupling is turned-on and the Rabi frequency has a general time dependence $\Omega(t)$, then it holds
\begin{equation}
\psi_j(t_0)=\phi_j(t_0)=\left( 
\begin{array}{c}
\Psi_{j}(t_0)\\
0
\end{array}
\right).
\label{t-t0}
\end{equation}
Due to the fact that Eq. (\ref{eqn:dnls}), which describes the evolution of $\phi$ for $t>t_0$, is 
identical to the one describing the evolution of $\psi$ for $t<t_0$ (i.e., without the Rabi coupling), 
it follows that for $t>t_0$ one has:
\begin{equation}
\phi_j(t)=\left( 
\begin{array}{c}
\Psi_{j}(t)\\
0
\end{array}
\right).
\label{t-larger}
\end{equation}
Using Eq. (\ref{unit:dnls}), one gets 
\begin{equation}
\psi_j(t)=\left( 
\begin{array}{c}
\cos{{\cal I}(t)} \cdot \Psi_{j}(t)\\
-i\sin{{\cal I}(t)} \cdot \Psi_{j}(t)
\end{array}
\right).
\label{psi-t-larger}
\end{equation}
Equation (\ref{psi-t-larger}) shows that the soliton can {\it tunnel} without losing its shape, and the 
effect of the transformation is 
just a change in the normalization. Hence, normalizing $\Psi_j$ to $1$ and denoting 
\begin{equation}
N_\alpha=\sum_j \mid \psi_{(\alpha)j}\mid^2
\label{number}
\end{equation}
the fraction of the number of particles in condensate $\alpha$, one gets $N_1=1$ and $N_2=0$ for $t \leq t_0$, as well as 
\begin{equation}
N_1=\cos^2{{\cal I}(t)},
\label{number-1}
\end{equation}
and $N_2=1-N_1$ for $t>t_0$. We notice that the result (\ref{number-1}) does not depend on the particular soliton solution 
chosen, nor on the momentum of the soliton.

\subsection{$1$D Chain}
 
\begin{figure}[t]
\begin{center}
\includegraphics[width=6.cm,height=8.cm,angle=270,clip]{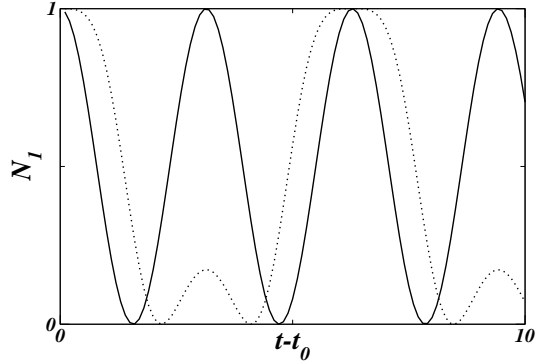} 
\caption{Fraction of the number of particles $N_1$ in the condensate $1$ 
as a function of time for $\Omega(t)=\Omega_0$ (solid line) 
and $\Omega(t)=\Omega_0 \sin{\beta(t-t_0)}$ (dotted line), where $\beta=\Omega_0/\hbar$.
In both cases, time is in units of $\Omega_0/\hbar$. 
}
\label{numbers}
\end{center}
\end{figure}

As an example of the previous results we considered the $1$D case, and in Fig.
\ref{numbers} we plot (for $t>t_0$) the number of particles $N_1$ in the first component, 
for a constant Rabi frequency $\Omega(t)=\Omega_0$ and for a sinusoidal Rabi frequency 
$\Omega(t)=\Omega_0 \sin{\beta(t-t_0)}$ versus time. One can clearly see that in both cases $N_1$ oscillates between the values 
one and zero, as 
expected.

As a further application of the previous analysis we consider the $1$D homogeneous 
problem with $V_j=0$. 
In this case, measuring time in units of $\hbar/2K_1$ and energies in units of $2K_1$, 
the rescaled coupled DNLSEs read
\begin{eqnarray}
i \frac{\partial \psi_{(1)j}}{\partial t}=- \frac{1}{2} \left( \psi_{(1)j+1}+\psi_{(1)j-1} \right)   +
\Lambda \left( \vert \psi_{(1)j}\vert ^{2} + \vert \psi_{(2)j}\vert ^{2} \right) 
\psi_{(1)j} +\omega(t) \psi_{(2)j},
\label{lceqa-1d-ad} \\
i \frac{\partial \psi_{(2)j}}{\partial t}=- \frac{1}{2} R \left( \psi_{(2)j+1}+\psi_{(2)j-1} \right)   +
\Lambda \left( \vert \psi_{(1)j}\vert ^{2} + \vert \psi_{(2)j}\vert ^{2} \right) 
\psi_{(2)j} +\omega(t) \psi_{(1)j},
\label{lceqb-1d-ad}
\end{eqnarray}
where $\Lambda=u/2K_1$, $\omega=\Omega/2K_1$ and $R=K_2/K_1$ is assumed to be equal to 1. 
At $t=0$ all particles are supposed to be in 
the first component in a soliton-like wavepacket and $\omega=0$ for $t<t_0$. 
It is known that on a chain, the single-component DNLSE soliton-like solutions 
can propagate for a long time even if the equation is not integrable \cite{hennig99}. 
Therefore, we consider, at $t=0$, a gaussian wavepacket centered 
at $\xi(t=0) \equiv \xi_0$, with initial momentum $p_0$ and width $\gamma(t=0)=\gamma_0 \equiv \sqrt{\alpha_0}$. 
Then, its temporal evolution can be analyzed by a variational approach, 
where we assume that the wavefunction has the following form
\begin{equation}
\label{gaussol}
\Psi_{n}^V(t)=\sqrt{{\cal K}} \, \cdot \, 
\exp\left[ -\frac{(n-\xi)^2}{\alpha} + ip(n-\xi) +
i \frac{\delta}{2}(n-\xi)^2\right],
\end{equation}
where $\xi(t)$ and $\gamma(t)=\sqrt{\alpha(t)}$ are the center and the width of the density  respectively, 
having conjugate momenta $p(t)$ and $\delta(t)$ 
(${\cal K}$ is a normalization factor).
Writing the equations of motion for $\xi,p,\alpha,\delta$ one obtains $p(t)=p_0$ and 
$\dot{\xi}=\sin{p_0} e^{-\eta}$, with $\eta=1/\left( 2\alpha \right) +\alpha \delta^2/8$. Imposing the conditions 
$\dot{\gamma}=0$ and $\dot{\delta}=0$, it follows that for $\cos{p_0}<0$ the value of $\Lambda$
(when $\gamma_0 \gg 1$) for which a (variational) soliton solution exists is \cite{trombettoni01}:
\begin{equation}
\label{lambda_sol}
\Lambda_{sol} = 2 \sqrt{\pi} \frac{\mid \cos{p_0} \mid} 
{\gamma_0}.
\end{equation}
The stability of the variational solutions for large times has been investigated numerically 
in \cite{malomed96,aceves96}. For the value $\Lambda_{sol}$ given by Eq. (\ref{lambda_sol}) one 
has $\alpha(t)=\alpha_0$, $\delta(t)=0$ and 
$\dot{\xi}=\sin{p_0} \cdot e^{-1/\left( 2\alpha_0 \right) } \approx \sin{p_0}$. Notice that $p=\pi$ corresponds to a gap soliton. 

\begin{figure}[t]
\begin{center}
\includegraphics[width=6.cm,height=8.cm,angle=270,clip]{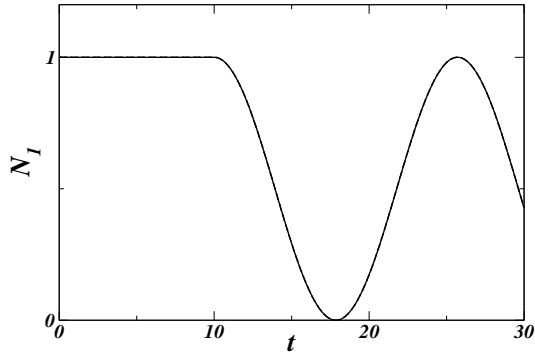} 
\caption{Fraction of the number of particles $N_1$ in the condensate $1$ 
as a function of time 
for $p=\pi$ (solid line) and 
$p=3\pi/4$ (dashed line) obtained from the numerical solution of the coupled DNLSEs 
(\ref{lceqa-1d-ad})-(\ref{lceqb-1d-ad}) with the initial condition given by (\ref{gaussol}) and $\Lambda=\Lambda_{sol}$. 
The two lines are indistinguishable and coincide 
with (\ref{number-1}), also plotted as a dotted line. We used the parameter values: $\Lambda=1$, $t_0=10$, 
$\omega=0.2$, $\gamma_0=50$.}
\label{numer}
\end{center}
\end{figure}

If at $t=t_0$ we turn on the Rabi switch with frequency $\omega(t)$, then we obtain
\begin{equation}
\psi_j(t)=\left( 
\begin{array}{c}
\cos{{\cal I}(t)} \cdot \Psi_{j}^V(t)\\
-i\sin{{\cal I}(t)} \cdot \Psi_{j}^V(t)
\end{array}
\right).
\label{psi-t-larger-1D}
\end{equation}
The numerical solution of the coupled DNLSEs (\ref{lceqa-1d-ad})-(\ref{lceqb-1d-ad}), with the initial condition 
$\psi_{(1)j}(t=0)=\Psi_j^V(t=0)$ given by (\ref{gaussol}) and $\Lambda=\Lambda_{sol}$, turns out to be in 
excellent agreement with (\ref{psi-t-larger-1D}) for different values of $p$, as shown in Fig.~\ref{numer} for $\omega$ 
constant.

\section{Different Tunneling Rates}

When the tunneling rates are different (i.e., $R=K_1/K_2 \neq 1$), the unitary transformation (\ref{unit:dnls}) does not really 
simplify the problem. Actually, Eqs. (\ref{lceqa-gen})-(\ref{lceqb-gen}) can be written in the compact form 
(\ref{eqn:stark:dnls}) with the notation
\begin{equation}
\label{not}
{\cal T} \psi_j \equiv {\cal T}_{K_1 ; K_2} \psi_j= 
- \left( 
\begin{array}{c}
K_1 \sum_{n(j)}\psi_{(1)n}\\
K_2 \sum_{n(j)}\psi_{(2)n}.
\end{array}
\right)
\end{equation}
By performing the transformation (\ref{unit:dnls}) one finally gets
\begin{equation}
i \hbar \frac{\partial {\phi_j}}{\partial t}  =
{\cal T}_{\tilde{K}_1 ; \tilde{K}_2} \phi_j+({\phi_j}^\dagger G \phi_j)
\phi_j+V_j \phi_{j} - \sigma_y \Delta {\cal S}(t) {\cal C}(t) \sum_{n(j)} \phi_j,  
\label{app-diff}
\end{equation}
where $\Delta=K_1-K_2$, ${\cal S}=\sin{{\cal I}}$, ${\cal C}=\cos{{\cal I}}$, $\tilde{K}_1(t)=K_1-\Delta {\cal S}^2(t)$, 
$\tilde{K}_2(t)=K_2+\Delta {\cal S}^2(t)$, and 
$\sigma_y=\left(
\begin{array}{cc}
0&-i\\
i&0
\end{array}
\right)$. 
It therefore follows that the effect of the transformation (\ref{unit:dnls}) is to replace the general time 
dependence $\Omega(t)$ of the Rabi switch, by the last term of Eq. (\ref{app-diff}). 
So, whenever the tunneling rates are different we study  
Eqs. (\ref{lceqa-gen})-(\ref{lceqb-gen}) directly.

\begin{figure}[t]
\begin{center}
\includegraphics[width=6.cm,height=8.cm,angle=270,clip]{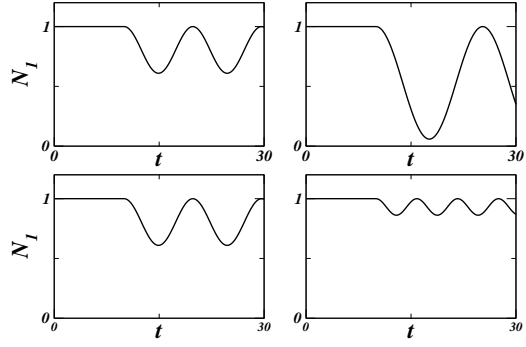} 
\caption{Fraction of the number of particles $N_1$ in the first component as a function of time for $p=\pi$ 
obtained from the numerical solution of the coupled DNLSEs for four different values of $R$, namely $0.5$ (top-left), 
$0.9$ (top-right), $1.5$ (bottom-left), and $2.0$ (bottom-right); 
the rest of the parameters are the same as in Fig. \ref{numer}.}
\label{erre}
\end{center}
\end{figure}

In Fig. \ref{erre} we plot the time dependence of $N_1$ for different values of $R$. 
It is observed that  
$N_1$ does not reach the value zero, i.e., the soliton cannot be  totally transferred to the other component. 
Moreover, it is clear that the larger the deviation of $R$ from $1$ is, the larger the minimum reached value of $N_1$ 
becomes, and therefore the more
inefficient the transferring process becomes.

These numerical results can be understood by means of a simple two-mode model, where for $t>t_0$ 
the variational wavefunction is assumed to be of the form
\begin{equation}
\psi_j^V=
\left( 
\begin{array}{c}
\psi_{(1)j}^V\\
\psi_{(2)j}^V
\end{array}
\right)=\Psi_j^V \cdot \left( 
\begin{array}{c}
\sqrt{N_1(t)} e^{i \varphi_1(t)}\\
\sqrt{N_2(t)} e^{i \varphi_2(t)}
\end{array}
\right).
\label{variational}
\end{equation}
Here, $\Psi_j^V$ is given by (\ref{gaussol}) and the variational parameters are now the center $\xi$, the momentum $p$, 
the width $\sqrt{\alpha}$ and its momentum $\delta$, plus $N_1$, $N_2$ (the number of particles in the two components) 
and the phases $\varphi_1$, $\varphi_2$. Note that a similar to Eq. (\ref{variational}) variational ansatz 
has been used in the past to study 
soliton dynamics in two linearly coupled continuous NLS equations describing pulse propagation in dual-core optical fibers \cite{var}.

The Lagrangian ${\cal L}$ is given by
\begin{equation}
{\cal L}=i \sum_j \psi_j^{V \dagger} \frac{\partial}{\partial t}\psi_j^V - {\cal H},
\label{LAG}
\end{equation}
where the Hamiltonian ${\cal H}$ is
$$
{\cal H}=-\frac{1}{2} \sum_j \left( \psi_{(1)j} \psi_{(1)j+1}^{\ast} + c.c. \right) 
-\frac{R}{2} \sum_j \left( \psi_{(2)j} \psi_{(2)j+1}^{\ast} + c.c. \right)
$$
\begin{equation}
+\omega \sum_j \left( \psi_{(1)j} \psi_{(2)j}^{\ast} + c.c. \right)+\frac{\Lambda}{2} \sum_j 
\left( \mid \psi_{(1)j} \mid^4 + \mid \psi_{(2)j} \mid^4 + 2 \mid \psi_{(1)j} \mid^2 \mid \psi_{(2)j} \mid^2 \right).
\label{Hamil}
\end{equation}

Omitting the details, the equations of motion for $p$ and $\xi$ are found to be 
\begin{eqnarray}
\dot{p} & = & 0 \label{var1}, \\
\dot{\xi} & = & \left( N_1 + R N_2 \right) \sin{p} \, \cdot \, e^{-\eta},  
\label{var2}
\end{eqnarray} 
(where $\eta=1/\left( 2\alpha \right)+\alpha \delta^2/8$), and those for $\alpha$ and $\delta$ are 
\begin{eqnarray}
\dot{\delta} & = & \left( N_1 + R N_2 \right) \cos{p} \cdot \Big(\frac{4}{\alpha^2}-\delta^2 \Big) 
e^{-\eta} + 2 \Lambda / \sqrt{\pi} \gamma^3, \label{var3} \\ 
\dot{\alpha} & = & 2 \left( N_1 + R N_2 \right) \alpha \delta \cos{p} \, \cdot \, 
e^{-\eta}.
\label{var4}
\end{eqnarray} 

One sees that for $R=1$, since $N_1+N_2=1$, Eqs. (\ref{var1})-(\ref{var4}) do not depend on the equations for 
$N_1$, $N_2$, $\varphi_1$, $\varphi_2$. When $R \neq 1$, the dynamics of the internal degrees of freedom is 
coupled with the phase-number dynamics. The equations for $N_1$ and $N_2$ are 
\begin{eqnarray}
\dot{N_1} & = & 2 \omega \sqrt{N_1 N_2} \sin{\varphi}\label{var5}, \\
\dot{N_2} & = & - 2 \omega \sqrt{N_1 N_2} \sin{\varphi}\label{var6}, 
\end{eqnarray}  
where the relative phase is given by 
\begin{equation}
\varphi=\varphi_1-\varphi_2,
\label{phase}
\end{equation}
and 
evolves in time according to
\begin{equation}
\dot{\varphi} = \left( 1-R\right) \cos{p} \, \cdot \, 
e^{-\eta}- \omega \left( \sqrt{\frac{N_2}{N_1}} - \sqrt{\frac{N_1}{N_2}} \right) \cos{\varphi}. 
\label{var7}
\end{equation}

To provide an estimate for the behaviour of $N_{1,2}$ from the system (\ref{var1})-(\ref{var7}), we introduce the 
approximation $e^{-1/\left( 2\alpha \right) } \approx 1$, which is reasonable for 
broad solitons. 
In this case, by introducing the fractional imbalance
\begin{equation}
z=N_1-N_2,
\label{imb}
\end{equation}
the following equations for $z$ and $\varphi$ are obtained
\begin{eqnarray}
\dot{z} & = & - 2 \omega \sqrt{1-z^2} \sin{\varphi}\label{var-zeta}, \\
\dot{\varphi} & = & \left( 1-R\right) \cos{p} + 2 \omega \frac{z}{\sqrt{1-z^2}} \cos{\varphi}\label{var-phi}. 
\end{eqnarray}  

\begin{figure}[t]
\begin{center}
\includegraphics[width=6.cm,height=8.cm,angle=270,clip]{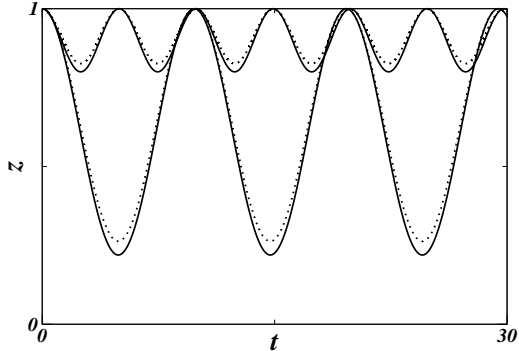} 
\caption{Fractional imbalance $z$ as a function of time for $p=\pi$ 
obtained from the numerical solution (solid line) of the coupled DNLSEs for two different values of $R$, 
namely $1.5$ (top), and $2.2$ (bottom); 
the dotted line is the solution of Eqs. (\ref{var-zeta})-(\ref{var-phi}) for the same values of $R$, 
namely $1.5$ (top), and $2.2$ (bottom). The values of the other parameters are the ones used in Fig.\ref{numer}.}
\label{comp}
\end{center}
\end{figure}

One readily realizes that Eqs. (\ref{var-zeta})-(\ref{var-phi}) are the equations of a dimer \cite{kenkre86,kenkre87}, 
but without the mass term (i.e., corresponding then to a non-interacting dimer) and with a detuning term. 
These equations describe the dynamics of a BEC in a double-well potential 
\cite{smerzi97,raghavan99} and analytical expressions for the quantities of interest 
can be found \cite{kenkre86,kenkre87,raghavan99}. One of the effects of the detuning is to 
induce the deviations from the complete transfer of the particles 
from one mode to the other, which is what is numerically found. In Fig.
\ref{comp}, we compare the numerical results 
obtained from the coupled DNLSEs with the solutions of Eqs. (\ref{var-zeta})-(\ref{var-phi}); 
it is clear that the agreement between the two is fairly good.

\section{Linearly coupled Ablowitz-Ladik equations}

The Ablowitz-Ladik equation is an integrable variant of the DNLSE \cite{ablowitz76}, which in 
dimensionless units can be expressed as 
\begin{equation}
\label{AL}
i  \frac{\partial \Psi_j}{\partial t} = - (\Psi_{j-1} + \Psi_{j+1}) (1+ \Lambda \mid \Psi_j \mid ^2).
\end{equation}
Accordingly, a system of two linearly coupled Ablowitz-Ladik equations can be written as
\begin{equation}
i \frac{\partial}{\partial t} \left( 
\begin{array}{c}
\psi_{(1)j}\\
\psi_{(2)j}
\end{array}
\right)=-\left( 
\begin{array}{c}
\psi_{(1)j+1}+\psi_{(1)j-1}\\
\psi_{(2)j+1}+\psi_{(2)j-1}
\end{array}
\right) \left[ 1+ \Lambda \left( \mid \psi_{(1)j} \mid ^2 + \mid \psi_{(2)j} \mid ^2 \right) \right] +
\omega(t) \left( 
\begin{array}{c}
\psi_{(2)j}\\
\psi_{(1)j}
\end{array}
\right).
\label{eqn:al}
\end{equation}
Notice that this version of the vector Ablowitz-Ladik equation
for $\omega=0$ constitutes the integrable vector (Manakov-like) generalization
of the one-component Ablowitz-Ladik model \cite{ablowitz04}. However,
it is different from other coupled models of this type with
either nonlinear \cite{bulow} or linear \cite{jianke} coupling.

By defining $\phi_i$ through the unitary transformation (\ref{unit:dnls}), substituting 
in Eq. (\ref{eqn:al}) and multiplying by $U^\dagger$, one obtains
\begin{equation}
i \frac{\partial}{\partial t} \left( 
\begin{array}{c}
\phi_{(1)j}\\
\phi_{(2)j}
\end{array}
\right)=-\left( 
\begin{array}{c}
\phi_{(1)j+1}+\phi_{(1)j-1}\\
\phi_{(2)j+1}+\phi_{(2)j-1}
\end{array}
\right) \left[ 1+ \Lambda \left( \mid \phi_{(1)j} \mid ^2 + \mid \phi_{(2)j} \mid ^2 \right) \right],
\label{eqn:al:phi}
\end{equation}
which is in fact the same as 
Eq. (\ref{eqn:al}) but without the Rabi term. In obtaining Eq. (\ref{eqn:al:phi}), one takes advantage of the fact 
that $\mid \psi_{(1)j} \mid ^2 + \mid \psi_{(2)j} \mid ^2 = \mid \phi_{(1)j} \mid ^2 + \mid \phi_{(2)j} \mid ^2 $.

Since the integrable Eq. (\ref{eqn:al:phi}) (for which soliton solutions
are analytically available \cite{ablowitz04}) 
does not depend on $\omega$, the analysis 
presented in Section II is applicable and yields 
the same results for the dynamics of an (exact) soliton wavefunction that corresponds to all particles being 
initially  in one of the condensates.

\vspace{1.cm}

\section{Conclusions}

In this work we considered the soliton dynamics in a system of two 
discrete, 
linearly-coupled, nonlinear Schr\"odinger equations. 
We showed that there is a unitary transformation that can be applied to the system and has the effect of
eliminating the time-dependent linear coupling terms in the case when the nonlinear coupling coefficients are
equal. We showed that the solitonic solutions that describe the number of particles of a two-component Bose gas, 
can oscillate from one species to the other. In the case where the 
tunneling rates are different, 
although the transformation can still be made, it has no simplifying 
effect on the analysis. 
Hence, we resorted to a numerical study of the problem and found good 
agreement with the results obtained from an effective two-mode model. 
These results indicate that the efficiency of the transfer mechanism
is reduced monotonically with the deviation from the equal tunneling rate 
limit (see also \cite{cont}). Finally, we showed that the 
same unitary transformation 
can also be applied in the analysis of a system of two 
linearly-coupled Ablowitz-Ladik equations,  
transforming the linearly coupled vector model into the
well-known integrable vector Ablowitz-Ladik equation.

There are 
many extensions that can be considered in connection
with this work. The same considerations can be applied to higher
dimensions, where it is known that while continuum nonlinear Schr{\"o}dinger
solitons are unstable to collapse, discrete ones may be stable
for sufficiently weak tunneling \cite{kevrekidis01,pre}.
On the other hand, one can generalize the phenomena examined
herein to the case of true spinor condensates, e.g. for spin-1 (or higher)
bosonic systems of $^{87}$Rb or $^{23}$Na which have been
experimentally realized \cite{higbie,chapman} and are
under intense theoretical investigation \cite{saito,you,kivshar,us}.
Notice that using a three-mode approach for analyzing the
transfer would be a particularly relevant \cite{chapman} approach
in that setting.

\vspace{0.5cm}

{\em Acknowledgements} 

A.T gratefully acknowledges discussions with P. Sodano, A. Smerzi, L. De Sarlo, C. Fort and F. Minardi and support 
by the MIUR project ``Quantum Noise in Mesoscopic Systems''. 
P.G.K. gratefully acknowledges support from 
NSF-CAREER, NSF-DMS-0505663 and NSF-DMS-0619492. The work of D.J.F. was partially supported by 
the Special Research Account of the University of Athens.



\vspace{0.5cm}


\begin{thebibliography}{99}

\bibitem{flach98}
S. Flach and C.R. Willis, Phys. Rep. {\bf 295}, 181 (1998).

\bibitem{scott99}
A. C. Scott, {\em Nonlinear Science: Emergence and
Dynamics of Coherent Structures} (Oxford University Press, Oxford, 1999).

\bibitem{hennig99}
D. Hennig and G. P. Tsironis, Phys. Rep. {\bf 307}, 333 (1999).

\bibitem{peli1} J. Hudock, P. G. Kevrekidis, B. A. Malomed, and D. N. Christodoulides, 
Phys. Rev. E {\bf 67}, 056618 (2003).

\bibitem{peli2} P. G. Kevrekidis and D. E. Pelinovsky, 
Proc. Roy. Soc. A {\bf 462}, 2671 (2006).


\bibitem{binary1} C. J. Myatt, E. A. Burt, R. W. Ghrist, E. A. Cornell, and C. E. Wieman, 
Phys. Rev. Lett. \textbf{78}, 586 (1997).

\bibitem{binary2} D. S. Hall, M. R. Matthews, J. R. Ensher, C. E. Wieman, and E. A. Cornell, 
Phys. Rev. Lett. \textbf{81}, 1539 (1998).

\bibitem{mertes07} K. M. Mertes, J. W. Merrill, R. Carretero-Gonz\'{a}lez, D. J. Frantzeskakis, P. G. Kevrekidis, and D. S. Hall, 
Phys. Rev. Lett. {\bf 99}, 190402 (2007).

\bibitem{spinor} D. M. Stamper-Kurn, 
M. R. Andrews, A. P. Chikkatur, S. Inouye, H.-J. Miesner, J. Stenger, and W. Ketterle, 
Phys. Rev. Lett. \textbf{80}, 2027 (1998); 
M.-S. Chang, 
C. D. Hamley, M. D. Barrett, J. A. Sauer, K. M. Fortier, W. Zhang, L. You, and M. S. Chapman, 
Phys. Rev. Lett. \textbf{92}, 140403 (2004).

\bibitem{KRb} G. Modugno, G.~Ferrari, G.~Roati, R.J.~Brecha, A.~Simoni, and M.~Inguscio, 
Science {\bf 294}, 1320 (2001).

\bibitem{LiCs} M. Mudrich, S. Kraft, K. Singer, R. Grimm, A. Mosk, and M. Weidem\"{u}ller, 
Phys.Rev. Lett. {\bf 88}, 253001 (2002).

\bibitem{minardi07} J. Catani, L. De Sarlo, G. Barontini, F. Minardi, and M. Inguscio, 
Phys. Rev. A {\bf 77}, 011603(R) (2008); 
G. Thalhammer, G. Barontini, L. De Sarlo, J. Catani, F. Minardi, and M. Inguscio, 
Phys. Rev. Lett. {\bf 100}, 210402 (2008).

\bibitem{morsch06} O. Morsch and M. Oberthaler, Rev. Mod. Phys. {\bf 78}, 179 (2006).

\bibitem{trombettoni01} A. Trombettoni and A. Smerzi, Phys. Rev. Lett. {\bf 86}, 2353 (2001).

\bibitem{kevrekidis01} 
P. G. Kevrekidis, K. \O. Rasmussen, and A. R. Bishop, Int. J. Mod. B {\bf 15},
2833 (2001).

\bibitem{smerzi07} A. Smerzi and A. Trombettoni, ''Optical Lattices: Theory'', in 
{\it Emergent Nonlinear Phenomena in Bose-Einstein Condensates: Theory and Experiment}, 
edited by P. G. Kevrekidis, D. J. Frantzeskakis, and R. Carretero-Gonz\'{a}lez, (Springer, Berlin, 2008), pp. 247-265.

\bibitem{ablowitz04} 
M. J. Ablowitz, B. Prinari, and A. D. Trubatch,
{\em Discrete and Continuous Nonlinear Schr\"odinger Systems}, 
(Cambridge University Press, Cambridge, 2004).

\bibitem{eisenberg98}
H. S. Eisenberg, Y. Silberberg, R. Morandotti, A. R. Boyd, and J. S. Aitchison, 
Phys. Rev. Lett. {\bf 81}, 3383 (1998).

\bibitem{malomed96} B. A. Malomed and M. I. Weinstein. Phys. Lett. A 
{\bf 220}, 91 (1996).

\bibitem{aceves96}
A. B. Aceves, 
C. De Angelis, T. Peschel, R. Muschall, F. Lederer, S. Trillo, and S. Wabnitz, 
Phys. Rev. E {\bf 53}, 1172 (1996).

\bibitem{duncan93} D. B. Duncan, J. C. Eilbeck, H. Feddersen, and J. A. D. Wattis, 
Physica D {\bf 68}, 1 (1993).

\bibitem{flach99} S. Flach and K. Kladko, Physica D {\bf 127}, 61 (1999).

\bibitem{gomez04} J. Gomez-Gardenes, 
L. M. Floria, M. Peyrard, and A. R. Bishop, 
Chaos {\bf 14}, 1130 (2004).

\bibitem{cuevas} T. R. Melvin, A. R. Champneys, P. G. Kevrekidis, and J. Cuevas, 
Phys. Rev. Lett. {\bf 97}, 124101 (2006).

\bibitem{oxtoby} O. F. Oxtoby and I. V. Barashenkov, 
Phys. Rev. E {\bf 76}, 036603 (2007).

\bibitem{mandel03} O.\ Mandel, M. Greiner, A. Widera, T. Rom, T. W. H\'{a}nsch, and I. Bloch, 
Phys. Rev. Lett. {\bf 91}, 010407 (2003).

\bibitem{ketterle} D. M.\ Stamper-Kurn and W. Ketterle, 
in {\it Coherent Atomic Matter Waves}, Les Houches Summer School Session LXXII, 
edited by R. Kaiser, C. Westbrook, and F. David 
(Springer, New York, 2001), pp. 137-217 (arXiv:cond-mat/0005001).

\bibitem{emergent} D. S.\ Hall, 
in {\it Emergent Nonlinear Phenomena in Bose-Einstein Condensates: Theory and Experiment}, 
edited by P. G. Kevrekidis, D. J. Frantzeskakis, and R. Carretero-Gonz\'{a}lez, (Springer, Berlin, 2008), 
pp. 307-327.

\bibitem{smerzi03} A. Smerzi, A. Trombettoni, T. Lopez-Arias, C. Fort, P. Maddaloni, F. Minardi, and M. Inguscio, 
Eur. Phys. J. B {\bf 31}, 457 (2003).

\bibitem{williams00} J.\ Williams, R. Walser, J. Cooper, E. A. Cornell, and M. Holland, 
Phys. Rev. A {\bf 61}, 033612 (2000).


\bibitem{nota} Notice that for a Bose-Bose mixture
composed by different bosonic species one has
$K_1 \neq K_2$, with $K_1$ larger than $K_2$ if $m_2 < m_1$ ($m_\alpha$ being the atomic mass of the bosonic species
$\alpha$); however, for a Bose-Bose mixture of different species, the Rabi tunneling can not be implemented.


\bibitem{decon} B. Deconinck, P. G. Kevrekidis, H. E. Nistazakis, and D. J. Frantzeskakis, 
Phys. Rev. A {\bf 70}, 063605 (2004).

\bibitem{trans}  M.V. Tratnik and J.E. Sipe, Phys. Rev. A {\bf 38}, 2011 (1988).

\bibitem{bradley} R. M. Bradley, B. Deconinck and J. N. Kutz, J. Phys. A {\bf 38}, 1901 (2005).

\bibitem{ablowitz76} M. J. Ablowitz and J. F. Ladik, J. Math. Phys. {\bf 17}, 1011 (1976).

\bibitem{manakov} S. V. Manakov, Zh. Eksp. Teor. Fiz. \textbf{65}, 505
(1973) [Sov. Phys. JETP \textbf{38}, 248 (1974)].

\bibitem{cont} H. Nistazakis, Z. Rapti, D. J. Frantzeskakis, P. G. Kevrekidis, P. Sodano, and A. Trombettoni,  
arXiv:0805.0189 (Phys. Rev. A, in press). 

\bibitem{var} C. Par\'e and M. Florja\'nczyk, Phys. Rev. A {\bf 41}, 6287 (1990); 
B. A. Malomed, I. M. Skinner, P. L. Chu, and G. D. Peng, 
Phys. Rev. E {\bf 53}, 4084 (1996);  A. I. Maimistov, Sov. J. Quantum Electron. {\bf 21}, 687 (1991). 

\bibitem{kenkre86} V. M. Kenkre and D. K. Campbell, Phys. Rev. B {\bf 34}, 4959 (1986).

\bibitem{kenkre87} V. M. Kenkre and G. P. Tsironis, Phys. Rev. B {\bf 35}, 1473 (1987).

\bibitem{smerzi97} A.\ Smerzi, S. Fantoni, S. Giovanazzi, and S. R. Shenoy, 
\newblock Phys.\ Rev.\ Lett. {\bf 79}, 4950 (1997).
 
\bibitem{raghavan99} S. Raghavan, A. Smerzi, S. Fantoni, and S. R. Shenoy, 
Phys. Rev. A {\bf 59}, 620 (1999).


\bibitem{bulow} A. B{\"u}low, D. Hennig and H. Gabriel,
Phys. Rev. E {\bf 59}, 2380 (1999).

\bibitem{jianke} B.A. Malomed and J. Yang, Phys. Lett. A {\bf 302}, 163 (2002).

\bibitem{pre} P.G. Kevrekidis, K.{\O}. Rasmussen and A.R. Bishop,
Phys. Rev. E {\bf 61}, 2006 (2000).

\bibitem{higbie} J. M. Higbie, L. E. Sadler, S. Inouye, A. P. Chikkatur, 
S. R. Leslie, K. L. Moore, V. Savalli, and D. M. Stamper-Kurn, 
Phys. Rev. Lett. {\bf 95}, 050401 (2005).

\bibitem{chapman} M.-S. Chang, Q. S. Qin, W. X. Zhang, L. You, and M. S. Chapman, 
Nature Phys. {\bf 1}, 111 (2005).

\bibitem{saito} H. Saito and M. Ueda,
Phys. Rev. A {\bf 72}, 023610 (2005).

\bibitem{you} W. Zhang, \"{O}. E. M\"{u}stecaplioglu, and L. You, 
Phys. Rev. A {\bf 75}, 043601 (2007).

\bibitem{kivshar} B.J. Dabrowska-W{\"u}ster, E. A. Ostrovskaya, T. J. Alexander, and Yu. S. Kivshar, 
Phys. Rev. A {\bf 75}, 023617 (2007).

\bibitem{us} H. E. Nistazakis, D. J. Frantzeskakis, P. G. Kevrekidis, 
B. A. Malomed, R. Carretero-Gonz\'{a}lez, and A. R. Bishop, Phys. Rev. A 76, 063603 (2007); 
H. E. Nistazakis, D. J. Frantzeskakis, P. G. Kevrekidis, B. A. Malomed, and R. Carretero-Gonz\'{a}lez, 
Phys. Rev. A 77, 033612 (2008).  
\end{thebibliography}
\end{document}